\documentclass[aps,prl,showpacs,superscriptaddress,twocolumn,amssymb,amsfonts]{revtex4-1}

\usepackage{amsmath,bm}
\usepackage{graphics}
\usepackage{dsfont}
\usepackage{subfigure}		
\usepackage{epsfig}
\usepackage{hyperref}		
\usepackage{units}

\def\beq{\begin{equation}}
\def\eeq{\end{equation}}
\def\beqn{\begin{eqnarray}}
\def\eeqn{\end{eqnarray}}

\begin{document}
\title{Unbounded Growth of Entanglement in Models of Many-Body Localization}
\author{Jens H. Bardarson}
\affiliation{Department of Physics, University of California, Berkeley, CA 94720, USA}
\affiliation{Materials Science Division, Lawrence Berkeley National Laboratory, Berkeley, CA 94720, USA}
\author{Frank Pollmann}
\affiliation{Max Planck Institute for the Physics of Complex Systems, D-0118 Dresden, Germany}
\author{Joel E. Moore}
\affiliation{Department of Physics, University of California, Berkeley, CA 94720, USA}
\affiliation{Materials Science Division, Lawrence Berkeley National Laboratory, Berkeley, CA 94720, USA}

\begin{abstract}
An important and incompletely answered question is whether a closed quantum system of many interacting particles can be localized by disorder.  The time evolution of simple (unentangled) initial states is studied numerically for a system of interacting spinless fermions in one dimension described by the random-field XXZ Hamiltonian.  Interactions induce a dramatic change in the propagation of entanglement and a smaller change in the propagation of particles.  For even weak interactions, when the system is thought to be in a many-body localized phase, entanglement shows neither localized nor diffusive behavior but grows without limit in an infinite system: interactions act as a singular perturbation on the localized state with no interactions.  The significance for proposed atomic experiments is that local measurements will show a large but nonthermal entropy in the many-body localized state.  This entropy develops slowly (approximately logarithmically) over a diverging time scale as in glassy systems.
\end{abstract}
\maketitle
One of the most remarkable predictions of quantum mechanics is that an arbitrarily weak random potential is sufficient to localize all
energy eigenstates of a single particle moving in one dimension~\cite{Anderson:1958fz,Lee:1985hn}.  In experiments on electronic systems, observation of localization is limited to low temperatures
because the interaction of an electron with its environment results in a loss of quantum coherence and a crossover to classical transport.  Recent work has proposed that, if there are electron-electron interactions but the electronic system is isolated from other degrees of freedom (such as phonons),
there can be a ``many-body localization transition'' even in a one-dimensional system for which all the single-particle states are
localized~\cite{Basko:2006hh,Oganesyan:2007ex,Monthus:2010gd,Berkelbach:2010ib,Pal:2010gr,Canovi:2011jd}.

Two important developments may enable progress on many-body localization beyond past efforts using analytical perturbation theory.  The first is that numerical methods like matrix-product-state based methods and large scale exact diagonalizations  enable studies of some, not all, important quantities in large systems.  The second is that progress in creating atomic systems where interactions between particles are strong but the overall many-atom system is highly phase
coherent~\cite{Kinoshita:2006bg} suggests that this many-body localization transition may be observable in experiments~\cite{Aleiner:2010ir,inguscio:2012}.  Note that many-body localization is connected to the problem of thermalization in closed quantum systems as a localized system does not thermalize.

The goal of the present work is to show that the many-body localized phase differs qualitatively, even for weak interactions, from the conventional, noninteracting localized phase.  The evolution of two quantities studied, the entanglement entropy and particle number fluctuations, show logarithmically slow evolution more characteristic of a glassy phase; however, the long-term behavior of these quantities is quite different. The growth of the entanglement entropy  has previously been observed~\cite{DeChiara:2006fo,Znidaric:2008cr} to show roughly logarithmic evolution for smaller systems and stronger interactions.  We seek here to study this behavior systematically over a wide range of time scales (up to $t\approx 10^9 J_{\perp}^{-1}$), showing that the logarithmic growth begins for arbitrarily weak interactions. We show that the entanglement growth {\it does not saturate} in the thermodynamic limit, and obtain additional quantities that distinguish among possible mechanisms.  Further discussion of our conclusions appears after the model, methods, and numerical results are presented.

{\it Model system.} -- One-dimensional (1D) $s=\frac{1}{2}$ spin chains are a natural place to look for many-body localization~\cite{Oganesyan:2007ex} as they are equivalent to 1D spinless lattice fermions.  To start, consider the XX model with random $z$ directed magnetic fields so that the total magnetization $S^z$ is conserved:
\begin{equation}\label{chain}
H_0 = J_{\perp} \sum_i \left( S_i^x S_{i+1}^x + S_i^y S_{i+1}^y \right) + \sum_i h_i S_i^z.
\end{equation}
Here the fields $h_i$ are drawn independently from the interval $[-\eta,\eta]$.  The eigenstates are equivalent via the Jordan--Wigner transformation
to Slater determinants of free fermions with nearest-neighbor hopping and random on-site potentials; particle number in the fermionic representation is
related to $S^z$ in the spin representation, so the $z$ directed magnetic field is essentially a random chemical potential.  Now every single-fermion
state is localized by any $\eta > 0$, and the dynamics of this spin Hamiltonian are localized as well: a local disturbance at time $t=0$ propagates only to some finite distance (the localization length) as $t \rightarrow \infty$.
As an example, consider the evolution of a randomly chosen $S^z$ basis state.  The coupling $J_{\perp}$ allows ``particles'' (up spins) to move, and entanglement entropy to develop, between two subregions $A$ and $B$.  But the total amount of entanglement entropy generated remains finite as $t \rightarrow \infty$  (Fig. 1), and the fluctuations of particle number eventually saturate as well (see below).  The entanglement entropy for the pure state of the whole system is defined as the von Neumann entropy $S = - {\rm tr}\ \rho_A \log \rho_A =  - {\rm tr}\ \rho_B \log \rho_B$ of the reduced density matrix of either subsystem. We always form the two bipartitions  by dividing the system at the center bond. 

The type of evolution considered here can be viewed as a ``global quench'' in the language of Calabrese and Cardy~\cite{Calabrese:2006bg} as the initial state is the ground state of an artificial Hamiltonian with local fields.  Evolution from an initial product state with zero entanglement can be studied efficiently via time-dependent matrix product state methods until a time where the entanglement becomes too large for a fixed matrix dimension.  Since entanglement cannot increase purely by local operations within each subsystem, its growth results only from propagation across the subsystem boundary, even though there is no conserved current of entanglement.

The first question we seek to answer is whether there is any qualitatively different behavior of physical quantities when a small interaction
\beq
H_{\rm int} =  J_z \sum_i S_i^z S_{i+1}^z
\eeq
is added.
With Heisenberg couplings between the spins ($J_z = J_{\perp}$), the model is believed to have a dynamical transition as a function of the dimensionless
disorder strength $\eta/J_{z}$~\cite{Oganesyan:2007ex,Monthus:2010gd,Pal:2010gr}.  This transition is present in generic eigenstates of the system and
hence exists at infinite temperature at some nonzero $\eta$.  The spin conductivity, or equivalently particle conductivity after the Jordan-Wigner
transformation, is zero in the many-body localized phase and nonzero for small enough $\eta/J_{z}$.
However, with exact diagonalization the system size is so limited that it has not been possible to estimate the location in the thermodynamic limit of the transition of eigenstates or conductivities.

We find that entanglement growth shows a qualitative change in behavior at infinitesimal $J_z$.  Instead of the expected behavior that a small interaction strength leads to a small delay in saturation and a small increase in final entanglement, we find that the increase of entanglement continues to times orders of magnitude larger than the initial localization time in the $J_z = 0$ case (Fig. 1). This slow growth of entanglement is consistent with prior observations for shorter times and larger interactions $J_z = 0.5 J_{\perp}$ and $J_z = J_{\perp}$~\cite{DeChiara:2006fo,Znidaric:2008cr}, although the saturation behavior was unclear.  Note that observing a sudden effect of turning on interactions requires large systems, as a small change in the Hamiltonian applied to the same initial state will take a long time to affect the behavior significantly.  We next explain briefly the methods enabling large systems to be studied.

\begin{figure}[tb]
  \includegraphics[width=0.9\columnwidth]{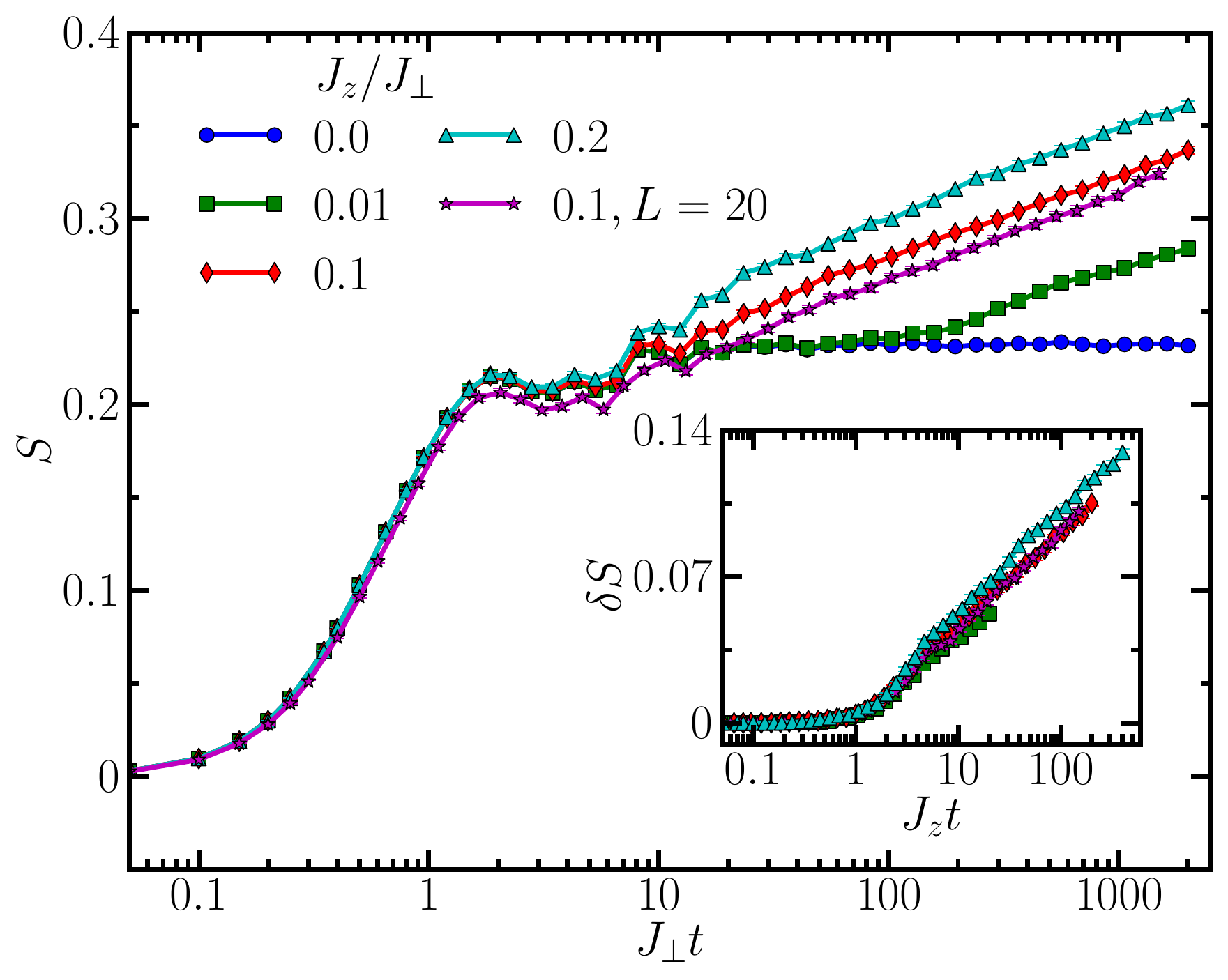} \\
  \includegraphics[width=0.9\columnwidth]{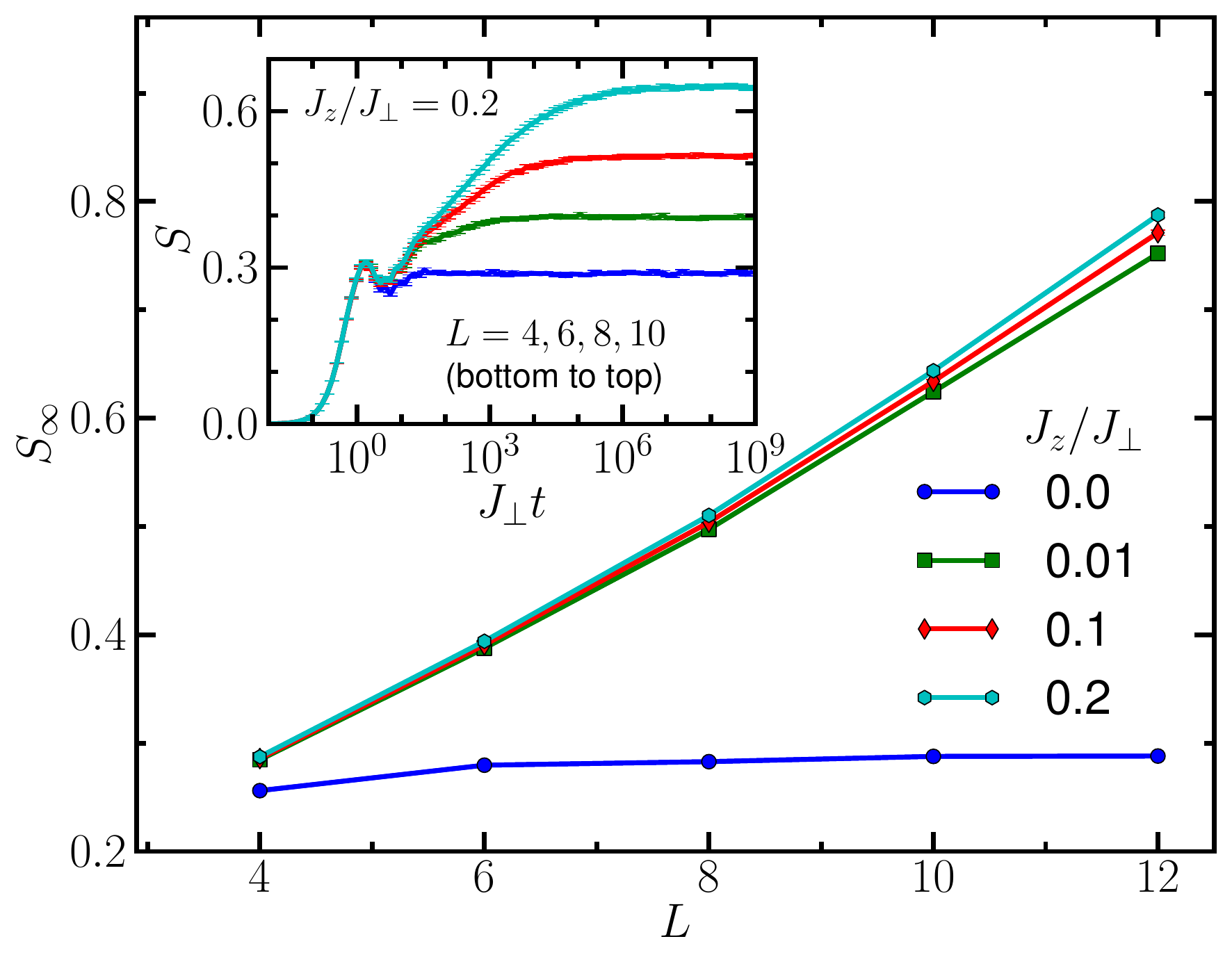}
  \caption{(a) Entanglement growth after a quench starting from  a site factorized $S^z$ eigenstate for different interaction strengths $J_z$ (we
	consider a bipartition into two half chains of equal size).  All data is for $\eta=5$ and  $L=10$, except for  $J_z=0.1$  where  $L=20$ is shown for
	comparison.The inset shows the same data but with a rescaled time axis and  subtracted $J_z=0$ values. (b) Saturation values of the entanglement
	entropy  as a function of $L$ for different interaction strengths $J_z$. The inset shows the approach to saturation.}
  \label{fig:S}
\end{figure}

{\it Numerical methodology.} -- To simulate the quench, we use the time evolving block decimation (TEBD)~\cite{Vidal:2003gb,Vidal:2007hx} method which provides an efficient method to
perform a time evolution of quantum states, $|\psi(t)\rangle=U(t)|\psi(0)\rangle$, in one-dimensional systems. The TEBD algorithm can be seen as a
descendant of the density matrix renormalization group~\cite{White:1992ie} method and is based on a matrix product state (MPS) representation~\cite{Fannes:1992dl,Ostlund:1995fx} of the wave functions.  We use a second-order Trotter decomposition of the short time propagator $U(\Delta t)=\exp(-i\Delta t H)$ into a product of term which acts only on two nearest-neighbor sites (two-site gates). After each application, the dimension of the MPS increases. To avoid an uncontrolled growth of the matrix dimensions, the MPS is truncated by keeping only the states which have the largest weight in a Schmidt decomposition.

In order to control the error, we check that the neglected weight after each step is small ($< 10^{-6}$). Algorithms of this type are efficient because they exploit the fact that the ground-state wave functions are only slightly entangled which allows for an efficient truncation. Generally the entanglement grows linearly as a function of time which allows only the simulation of rather short times. In our case, however, the entanglement growth is logarithmic and thus it is possible to perform a time evolution over long time scales \cite{Znidaric:2008cr}. 

In order to study the asymptotic ($t\rightarrow \infty$) behavior, we employ exact diagonalization techniques. We diagonalize the full Hamiltonian and
construct $U(t)$ which allows us to access very late times after the quench. Since the Hilbert space grows exponentially with the system size, we can
only consider rather small systems. All the data in this paper are obtained by averaging over more than $10^4$ field configurations $\{h_i\}$ starting
from a 
random product state $|\psi(0)\rangle=|m_1\rangle|m_2\rangle\dots |m_L\rangle$, where $|m_j\rangle \in \{|\uparrow\rangle$,$|\downarrow\rangle\}$ are eigenstates of the $S^z_j$
operators (for the saturation data we always start from a N{\'e}el state to reduce noise -- the data are qualitatively the same for a random initial state).

\begin{figure}[tb]
  \includegraphics[width=0.9\columnwidth]{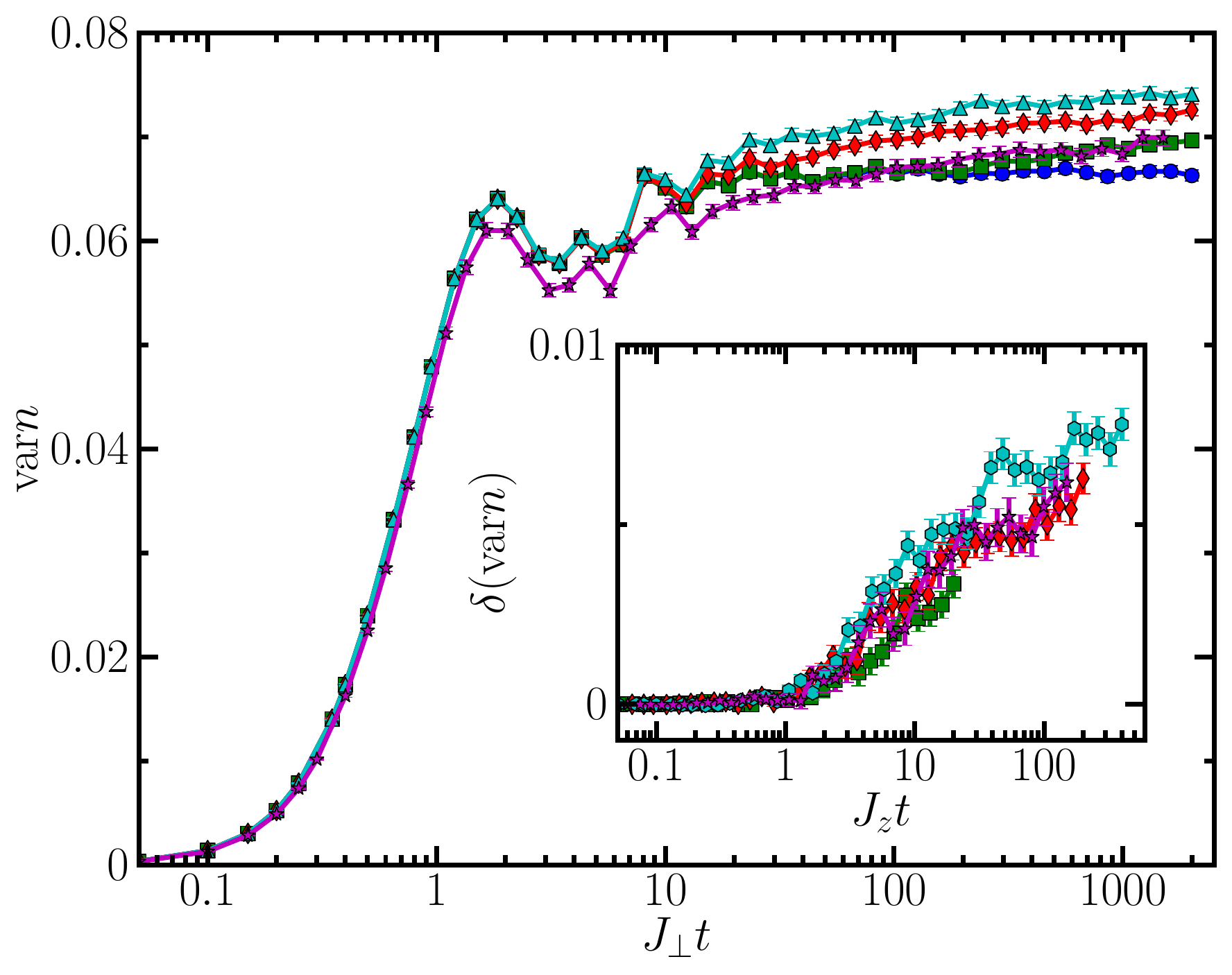} \\ 
  \includegraphics[width=0.9\columnwidth]{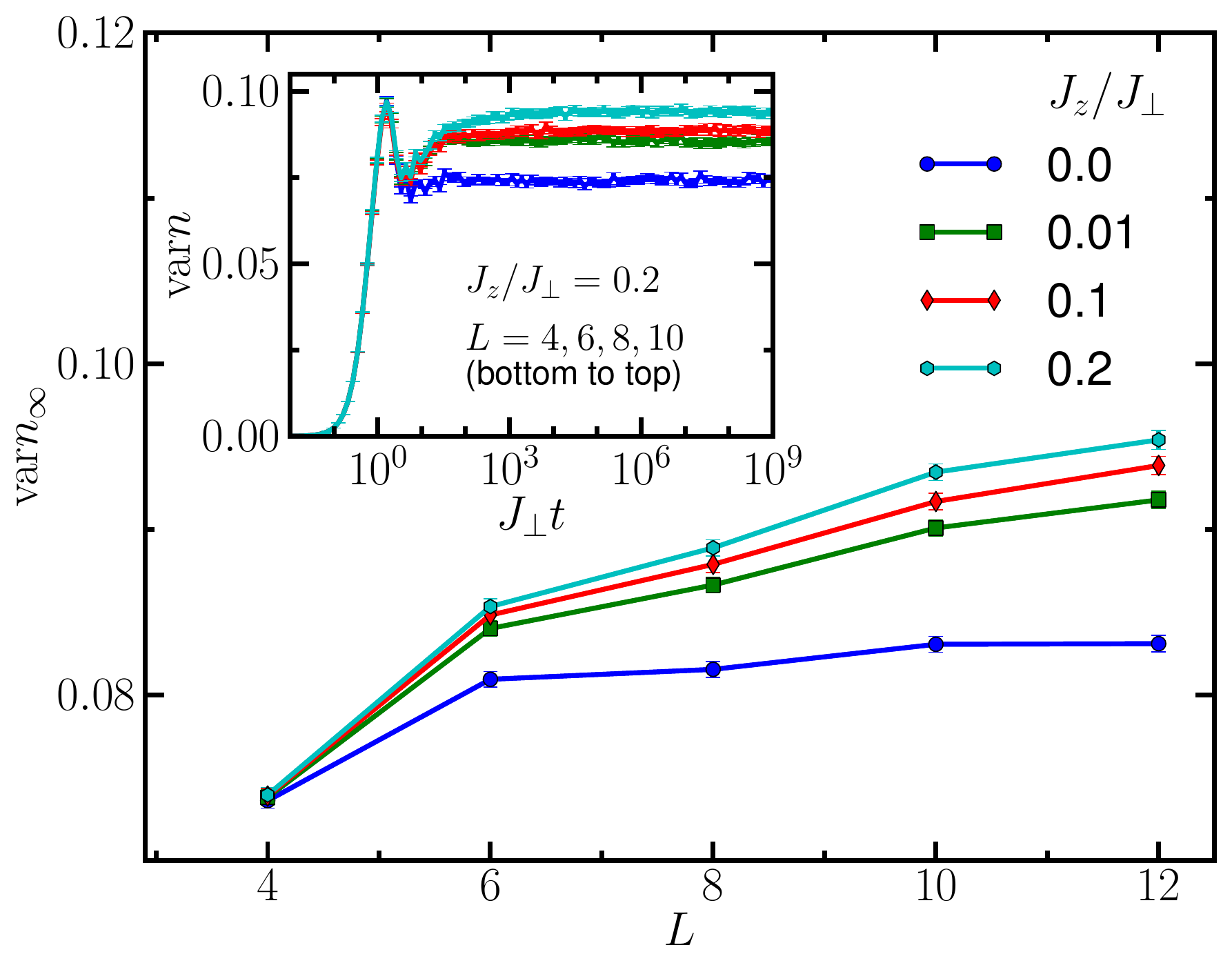}
  \caption{(a)  Growth of the particle number fluctuations of a half chain after a quench starting from  a site factorized $S^z$ eigenstate for different
	interaction strength $J_z$. All data  is for $\eta=5$ and $L=10$ except for  $J_z=0.1$  where  $L=20$ is shown for comparison. The inset shows the
	same data but with a rescaled time axis and  subtracted $J_z=0$ values. (b) Saturation values of the particle number fluctuations  as a function of $L$
	for different interaction strengths $J_z$. The inset shows the approach to saturation.}
  \label{fig:varn}
\end{figure}

{\it Saturation behavior and entropy.} -- In Fig.~\ref{fig:S} upper panel, we plot the time evolution of the half-chain entanglement entropy. The initial quick rise of the entanglement till time $J_\perp t
\sim 1$ corresponds to expansion of wave packets to a size of the order of the localization length. This rise is independent of the interaction strength
$J_z$. After this initial expansion, the entanglement saturates in the noninteracting case but increases logarithmically for any nonzero
interaction strength. The time at which the logarithmic growth starts is $J_z t \sim 1$, and all curves for different interacting strengths roughly collapse on a
single curve when plotted against $J_z t$ (see inset). In the case of $J_z/J_\perp = 0.1$ we
have plotted the entanglement for both $L = 10$ and $L = 20$. While absolute value of the entanglement differs slightly, the two curves are
indistinguishable after subtracting the noninteracting value (inset), suggesting that for the entanglement at the time scales explored, finite size
effects are small.

In an infinite system, we expect the logarithmic growth of the entanglement to continue indefinitely. In a finite system, in contrast, it will saturate
at late times at some value $S_\infty$. The lower panel in Fig.~\ref{fig:S} shows the approach to saturation and the system size dependence of this saturation value, as obtained by
exact diagonalization. In the noninteracting case, this value is independent of system size as expected if the localization length $\xi \ll L$. In the
presence of interaction the saturation value is independent of the interaction strength (we have checked that the small difference between the curves at
large $L$ goes away as the saturation time is increased) and follows a volume law. The time $t_{\rm sat}$ at which the entanglement entropy saturates is consistent with the scaling $\log t_{\rm sat} \sim L$.  For small sizes the result is roughly consistent with
Ref.~\onlinecite{DeChiara:2006fo}, although going to longer times we find that the saturation follows a volume law rather than being logarithmic in
subsystem size as conjectured there.  Such logarithmic dependence occurs at certain random-singlet quantum critical points in one
dimension~\cite{Refael:2004hu}, but the present system is not critical by standard definitions. 
Recent work observes a slower $\log \log t$ growth of entanglement in a quench to the random-singlet critical point of a random-field Ising chain~\cite{Igloi:2012in}.

The observation of a volume law for the saturation entanglement entropy suggests at first glance that the system may be partially thermalized, as a finite entanglement entropy per site is consistent with a thermal entropy.  However, the saturation value of the entanglement entropy per site is much lower than the infinite temperature thermal value associated with 2 states per site.  Since, as shown below, particle number propagates much less than entanglement and may well be localized, the entanglement entropy at saturation may result from an incomplete thermalization where the system locally equilibrates among a restricted ensemble with particular values of particle number, energy, and other conserved quantities.  This would be a many-body-localized version of the generalized Gibbs ensemble describing thermalization in some systems without disorder.

The time evolution of the half-chain particle number fluctuations, i.e., the variance of the total spin on half the chain, is plotted in the upper panel of Fig.~\ref{fig:varn}. The behavior of the particle number fluctuations is qualitatively different than the entanglement entropy. Interactions do enhance the particle number fluctuations, but while there are signs of a
logarithmic growth as for the entanglement, this growth slows down with time. The saturation values, similarly, are interaction dependent and
only fractionally larger than in the noninteracting case. This does not seem to be a case of the system not having reached saturation, as increasing the
saturation time by two orders of magnitude has a negligible effect on the saturation values. The half-chain energy variance behaves in a manner similar
to the particle variance (data not shown). It would therefore seem that particle transport is not fully ergodic,
consistent with a many-body localized phase.

{\it Conclusions.} -- Our results are consistent with a phase diagram consisting of a conventionally localized phase at zero interaction, a quasilocalized phase for moderate interactions, and a delocalized phase when interactions dominate disorder.  We use the term ``quasilocalized'' to describe zero conductivity (particles are either localized or subdiffusive) while other quantities such as entanglement show continued evolution for long times.  The entanglement entropy difference between Anderson and many-body localized phases will appear in any local measurement of entropy, since all local measurements on a subsystem are entirely determined by the reduced density matrix of that subsystem.~\footnote{Models  can be constructed using long-range interactions for a many-body localized phase with zero particle number and energy conductivities, but with infinite saturation entanglement; an example pointed out to the authors by V. Oganesyan and D. Huse is the evolution of an initial $\sigma^x$ product state in a longitudinal-field Ising Hamiltonian with exponentially decaying effective interaction.}

It is possible that the particle and energy variances are not actually saturating but increase without limit, which cannot
strictly be ruled out from our data.  This would still be consistent with zero conductivity if, for example, the continuing increase results from subdiffusive motion of particles; alternately there could be diffusion at unobservably long times and no many-body localized phase.  The quantities studied here by matrix-product-state methods are not directly comparable to quantities previously studied by exact diagonalization, e.g., the level statistics of eigenstates~\cite{DeChiara:2006fo,Oganesyan:2007ex} or correlations in a single eigenstate~\cite{Pal:2010gr}, and there could be a sharp transition as a function of $\eta$ in such quantities while the real-time evolution studied here is insensitive to that transition.  However, if this is the case, it means that experiments must be carefully designed to detect the transition as the dynamics of a simple initial state with strong disorder are characterized better by glassy dynamics than ordinary localization.

If the ultimate dynamics on long time scales is dephased because one part of the system sees the remainder as a ``bath'', then classical dynamics is appropriate, and in classical physics it is well known that quenched random systems need not be
diffusive, especially in 1D. One model for slow dynamics was introduced by Sinai~\cite{Sinai:1982ta}, who showed that 1D Brownian motion is modified by a static Gaussian-correlated random force $F$,
\beq
\frac{dx}{dt} = \nu(t) + F(x).
\eeq
The displacement $R = \sqrt{\langle x^2 \rangle}$ of a particle from its initial location scales asymptotically as $R \sim \log^2(t)$
 rather than $R \sim \sqrt{t}$ as in ordinary diffusion.  The scaling of $R$ can be obtained from a simple estimate of the time required to cross an
 occasional region of unfavorable potential~\cite{Fisher:1984dt}.  Another possibly relevant classical model follows from considering classical rate
 equations with a broad distribution of rates~\cite{Amir:2010by}, which leads to logarithmic diffusion $R \sim \log t$ for intermediate times before ultimately crossing over to diffusion.

In conclusion, the many-body localized phase of the random-field XXZ Hamiltonian is fundamentally different in at least one measurable aspect, the dynamics of entanglement, from noninteracting Anderson localization.  The entanglement increases slowly until a saturation time scale, which diverges in the infinite system.  Two important questions for future work are a full characterization of other dynamical properties of this phase, including whether particles and energy are fully localized or propagate subdiffusively, and an understanding of how these properties connect to the phase transition to a delocalized phase.  Our results imply that local measurements starting from some initial states will show a large but nonthermal entropy in the many-body localized state in the presence of interaction, which is of particular  relevance for experiments on atoms in disordered optical lattices \cite{SanchezPalencia:2010db, Pasienski:2010bd}.

The authors acknowledge conversations with P. Calabrese, D. Huse, V. Oganesyan, and G. Refael, and support from DOE BES DMSE (J. H. B) and ARO via the DARPA OLE program (J. E. M.).

\bibliography{mbl.bib}
\end{document}